\documentclass[conference]{IEEEtran}

\def\boxend{\hspace*{\fill} $\QED$}

\newtheorem{theorem}{Theorem}

\newtheorem{definition}{Definition}

\def\done{\hspace*{\fill} $\framebox[2mm]{}$}
\def\done{\boxend}

\usepackage{cite}
\usepackage{algorithm}
\usepackage{algorithmic}
\usepackage{array}

\usepackage{graphicx}

\begin{document}
\bibliographystyle{IEEEtran}
%
\title{Link Scheduling in Multi-Transmit-Receive Wireless Networks}


\author{\authorblockN{Hong-Ning Dai\\}
\authorblockA{Macau University of Science and Technology, Macau\\
hndai@ieee.org\\}
\and
\authorblockN{Soung Chang Liew and Liqun Fu\\}
\authorblockA{The Chinese University of Hong Kong, Hong Kong\\
\{soung,lqfu6\}@ie.cuhk.edu.hk\\}}

\maketitle

\begin{abstract}
This paper investigates the problem of link scheduling to meet traffic demands with minimum airtime in a multi-transmit-receive (MTR) wireless network. MTR networks are a new class of networks, in which each node can simultaneously transmit to a number of other nodes, or simultaneously receive from a number of other nodes. The MTR capability can be enabled by the use of multiple directional antennas or
multiple channels. Potentially, MTR can boost the network capacity significantly. However, link scheduling that makes full use of the MTR capability must be in place before this can happen. We show that optimal link scheduling can be formulated as a linear program (LP). However, the problem is NP-hard because we need to find all the maximal independent sets in a graph first before the LP can be set up. We propose two computationally efficient algorithms, called Heavy-Weight-First (HWF) and Max-Degree-First (MDF) to solve this problem. Simulation results show that both HWF and MDF can achieve superior performance in terms of runtime and optimality. Specifically, we have conducted 1,000 simulation experiments with different network topologies and traffic demands. On average, the HWF and MDF solutions are within 90\% of the optimal solutions.

\end{abstract}
\IEEEpeerreviewmaketitle

\section{Introduction}

This paper concerns the problem of link scheduling to minimize airtime usage in a new class of wireless networks called multi-transmit-receive (MTR) wireless networks. In an MTR network, a node can simultaneously transmit to a number of other nodes, or simultaneously receive from a number of other nodes. However, a node cannot simultaneously transmit and receive (i.e., the half-duplexity is still in place). Enabling this capability of MTR networks is the use of multiple directional antennas at a node \cite{Raman:hotnets04,Patra:NSDI07,Raman:Mobicom05,Surana:NSDI08,Nedevschi:Mobicom2008} or the use of multiple
channels on multiple collocated radios at a node \cite{Sandeep:mobicom2009}. Potentially, this capability can increase the network capacity significantly \cite{Raman:Mobicom05,Patra:NSDI07,Sandeep:mobicom2009}. Details about MTR networks will be presented in Section \ref{sec:model}.

\begin{figure}[t]
\centering
\includegraphics[width=4cm]{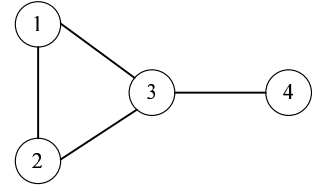}
\caption{A four-node network}
\label{fig:4node-ex}
\end{figure}

Potentially, MTR networks can schedule more wireless links than conventional wireless networks. Take Fig. \ref{fig:4node-ex} as an example. In this four-node network, the mutually connected nodes 1 and 2 are connected with nodes 3, which is in turn connected with node 4. Since there are four edges, there are totally eight directional links. Denote the link set by $\{(1,2),(1,3),(2,1),(2,3),(3,1),(3,2),(3,4),(4,3)\}$. For a conventional network, where multiple simultaneous transmissions or receptions at a node are not allowed, at most two links can be active at a given time (e.g., $(1,2)$ and $(4,3)$). However, an MTR network allows three links to be active simultaneously (e.g., $(1,2)$,$(1,3)$,$(4,3)$).

This paper considers the link-scheduling problem of determining the minimum Time Division Multiple Access (TDMA) frame length while fulfilling the traffic demands in MTR networks. Although there have been prior studies on MTR networks \cite{Raman:hotnets04,Raman:infocom06,Raman:Mobicom05,Subramanian:2006,Patra:NSDI07,Surana:NSDI08,Sheth:infocom07,Raman:www07,Nedevschi:Mobicom2008}, this particular link scheduling problem (which is a well studied classical problem under the context of non-MTR networks \cite{Das:infocom2005,liqun:icc2008,Borbash1:Ephremides,Tang:TVT2006,liqun:infocom2009,Borbash:Ephremides}) has not been investigated as far as we know. Previously proposed MTR MAC protocols, such as 2P \cite{Raman:Mobicom05}, WiLDNet \cite{Patra:NSDI07} and JazzyMAC \cite{Nedevschi:Mobicom2008}, are not efficient in that a node needs to maintain all of its links in transmit mode for the same time duration regardless of the actual link traffic demands. The links with lower traffic demands will sit idle while the other links of the node transmit; meanwhile, the nodes at the other ends of the idle links are not allowed to transmit - this is purely a constraint imposed by the MAC rather than an MTR constraint. Consider the four-node example in Fig. \ref{fig:4node-ex} again. Suppose a traffic demand for the above link set is $[1, 1, 1, 1, 1, 1, 2, 1]$. Then, 2P, WiLDNet and JazzyMAC obtain a sub-optimal schedule, $\{(1,2),(1,3),(4,3)\}$,$\{(2,1),(2,3)\}$,$\{(3,4)\}$,$\{(3,1),(3,2),\\(3,4)\}$, which requires four time slots. However, only three time slots are required for an optimal schedule: $\{(2,1),(3,1),(3,4)\}$, $\{(1,2),(3,2),(3,4)\}$,$\{(1,3),(2,3),\\(4,3)\}$. 

The primary research contributions of our paper are summarized as follows.
\begin{itemize}

\item[1.] We provide a formal specification of an MTR network, and formulate the link-scheduling problem of determining the minimum frame length required to meet the underlying link-traffic demands.

\item[2.] We show that solving the link scheduling problem optimally is NP-hard, since we need find all the maximal independent sets (MIS) in a graph.

\item[3.] We propose two computationally efficient heuristic algorithms to tackle this problem. The first algorithm is a Heavy-Weight-First (HWF) algorithm, which gives priority to the links with the heaviest traffic demands in its schedule. The second algorithm is a MAX-Degree-First (MDF) algorithm, which gives priority to the links with the maximum degree in a conflict graph in its schedule.

\item[4.] We conduct extensive simulations based on regular and random network topologies, with symmetric and asymmetric traffic demands. The simulation results show that both HWF and MDF can typically obtain solutions within 90\% of the optimal solutions over 1,000 simulation experiments based on symmetric and asymmetric traffic demands.

\end{itemize}

The rest of the paper is organized as follows. In Section \ref{sec:model}, we present the network model, basic assumptions and problem formulation. Section \ref{sec:alg} presents two heuristic algorithms. We show the simulation results in Section \ref{sec:sim}. Section \ref{sec:con} concludes the paper. 

\section{Network Model and Problem Formulation}
\label{sec:model}
In this section, we specify the MTR network model formally. We then formulate the link-scheduling problem of finding the minimum frame length in MTR networks as a linear program (LP). The previously proposed MTR MAC protocols (e.g., 2P \cite{Raman:Mobicom05}, WiLDNet \cite{Patra:NSDI07} and JazzyMAC \cite{Nedevschi:Mobicom2008}) are shown to be sub-optimal.

A Multi-Transmit-Receive (MTR) network is defined as follows.
\begin{definition}
\label{def:mtr}
In an MTR network, each node has a set of neighbor nodes
with whom it forms links. At any given time,

\begin{enumerate}
\item [R1.] A node can transmit simultaneously on a subset of its
outgoing links.

\item [R2.] A node can receive simultaneously on a subset of its 
incoming links.

\item [R3.] A node cannot do operations R1 and R2 simultaneously, (i.e., 
a node cannot transmit and receive simultaneously).

\end{enumerate}
\end{definition}

Given an MTR network, we are interested in how to minimize the TDMA slots required to meet the underlying link traffic demands.

\subsection{Centralized Scheduling Problem}
Let the link traffics be specified by the traffic matrix $T=[t_{ij}]$, where $t_{ij}$
is the amount of traffic from node $i$ to its neighboring
node $j$. At any given time, let the set of active links in the 
network be indicated by an indicator matrix, $M^{(k)}=[m^{(k)}_{ij}]$,
where $m^{(k)}_{ij}=1$ if link $(i,j)$ is active, and $m^{(k)}_{ij}=0$
if link $(i,j)$ is inactive.

\begin{definition}
An indicator matrix is called a \textit{matching matrix} if all nodes
conform to rules R1, R2 and R3.
\end{definition}

Let us consider the four-node network as shown in Fig. \ref{fig:4node-ex}. 
In this network, an example of a matching matrix is:
\begin{displaymath}
M^{(1)}=
\left(\begin{array}{cccc}
0 & 1 & 1 & 0 \\
0 & 0 & 0 & 0 \\
0 & 0 & 0 & 0 \\	
0 & 0 & 1 & 0
\end{array}\right)
\end{displaymath}
where matrix $M^{(1)}$ indicates that node 1 simultaneously 
transmits to nodes 2 and 3 while node 4 transmits to node 3.
Note that the matching matrix $M^{(1)}$ is \textit{maximal} in that
you cannot turn any of its $0$ elements to $1$ without 
violating R3.

\begin{definition}
A matching matrix is \textit{maximal} if none of its 0 elements can be turned to 1 (while maintaining all its 1 elements at 1) without violating the rules in definition 1. 
\end{definition}

When we consider the link scheduling, we only need to consider the 
maximal matching matrix. Suppose there be $K$ maximal matching matrices.
Then, the problem that we are considering is as follows:
\begin{eqnarray}
\label{eqn:central}
\min & \sum^{K}_{k=1}x_k \nonumber\\
s.t. & \sum^{K}_{k=1}M^{(k)}x_k \geq T \nonumber\\
& x_k \geq 0 \textrm{ for all } k
\end{eqnarray}
where $x_k$ denotes the number of time slots allocated to maximal matching matrix $M^{(k)}$.

\subsection{Sub-optimal Scheduling}
In previously proposed MAC protocols for MTR networks networks, such as 2P \cite{Raman:Mobicom05}, WiLDNet \cite{Patra:NSDI07} and JazzyMAC \cite{Nedevschi:Mobicom2008}, each node alternates between two phases (2P):  simultaneous reception (SynRx) and simultaneous transmission (SynTx). In addition, a node is required to maintain all of its links in transmit mode for the same time duration regardless of the link traffic demands, resulting in inefficiency. In particular, the simultaneous synchronized operations in these MAC protocols indicate that when a node transmits, none of its neighbor nodes can transmit. As far as scheduling is concerned, this constraint is equivalent to (virtually) turning R1 in Definition 1 to a more restrictive requirement, as follows:
\begin{enumerate}
\item [R1$'$] When a node transmits, it transmits on all its outgoing links.
\end{enumerate}

Constraint R1$'$ plus the half-duplexity in constraint R3 
implies that the neighbors of a node cannot
transmit at the same time.

Since with R1$'$, when a node $i$ transmits, it transmits on all outgoing links, 
we might as well replace the traffic requirements for outgoing traffic
from node $i$, $(t_{i1},t_{i2},...,t_{iN})$, by one single number, 
$t_i=\max_{j} t_{ij}$. Then $t=(t_i)$ is the traffic vector 
describing the transmission requirements of all nodes.

Let $S^{(l)}=(s_i^{(l)})$ be a column indicator vector in which $s_i^{(l)}=1$
if node $i$ transmits and $s_i^{(l)}=0$ if node $i$ does not transmit.  
With respect to the graph describing the network, $S^{(l)}$ is 
basically an independent set\footnote{An independent set is a subset
of vertices such that no edge joins any two of them.} 
if it is to conform to R1$'$, R2, and R3. 
It suffices to consider the \textit{maximal} independent set (MIS)\footnote{A maximal
independent set is an independent set that is not the subset of another
independent set.} 
in our scheduling problem. Suppose that there are $L$ MIS. Then, 
the scheduling problem can be formulated as follows:
\begin{eqnarray}
\label{eqn:sub-opt}
\min & \sum^{L}_{l=1}x_l \nonumber\\
s.t. & \sum^{L}_{l=1}S^{(l)}x_l \geq t \nonumber\\
& x_l \geq 0 \textrm{ for all } l
\end{eqnarray}

Since Eq. (\ref{eqn:sub-opt}) is defined in a more restrictive way,
the solution to Eq. (\ref{eqn:central}) cannot be worse than that of
Eq. (\ref{eqn:sub-opt}). 

Consider the four-node example (Fig. \ref{fig:4node-ex}) again. 
In addition, suppose we have the following traffic requirements:
\begin{displaymath}
T=
\left(\begin{array}{cccc}
0 & 1 & 1 & 0 \\
1 & 0 & 1 & 0 \\
1 & 1 & 0 & 2 \\	
0 & 0 & 1 & 0
\end{array}\right)
\end{displaymath}

Then, we replace the outgoing traffic requirements of each node by a single number
and get a vector:
\begin{eqnarray}
t=[1,1,2,1]^T\nonumber
\end{eqnarray}

In the four-node example, there are three maximal independent sets (MIS):
\begin{eqnarray}
S^{(1)}=[1,0,0,1]^T\nonumber\\
S^{(2)}=[0,1,0,1]^T\nonumber\\
S^{(3)}=[0,0,1,0]^T\nonumber
\end{eqnarray}

We can verify that the optimal solution to the problem defined in 
Eq. (2) is given by 
\begin{eqnarray}
x_1=1,&x_2=1,&x_3=2\nonumber
\end{eqnarray}

Thus, a total of four units of airtime is needed.

Now, let us go back to the original problem defined in Eq. (1). 
First of all, the maximal matching matrices corresponding to 
the above MIS are:

\begin{displaymath}
M^{(1)}=
\left(\begin{array}{cccc}
0 & 1 & 1 & 0 \\
0 & 0 & 0 & 0 \\
0 & 0 & 0 & 0 \\	
0 & 0 & 1 & 0
\end{array}\right)\,
\end{displaymath}
\begin{displaymath}
M^{(2)}=
\left(\begin{array}{cccc}
0 & 0 & 0 & 0 \\
1 & 0 & 1 & 0 \\
0 & 0 & 0 & 0 \\	
0 & 0 & 1 & 0
\end{array}\right)\,
\end{displaymath}
\begin{equation}
\label{eqn:eg3}
M^{(3)}=
\left(\begin{array}{cccc}
0 & 0 & 0 & 0 \\
1 & 0 & 0 & 0 \\
1 & 1 & 0 & 1 \\	
0 & 0 & 0 & 0
\end{array}\right)
\end{equation}

In addition, there are three additional maximal matching matrices, shown
as follows:

\begin{displaymath}
M^{(4)}=
\left(\begin{array}{cccc}
0 & 0 & 0 & 0 \\
1 & 0 & 0 & 0 \\
1 & 0 & 0 & 1 \\	
0 & 0 & 0 & 0
\end{array}\right)\,
\end{displaymath}
\begin{displaymath}
M^{(5)}=
\left(\begin{array}{cccc}
0 & 1 & 0 & 0 \\
0 & 0 & 0 & 0 \\
0 & 1 & 0 & 1 \\	
0 & 0 & 0 & 0
\end{array}\right)\,
\end{displaymath}
\begin{equation}
\label{eqn:eg4}
M^{(6)}=
\left(\begin{array}{cccc}
0 & 0 & 1 & 0 \\
0 & 0 & 1 & 0 \\
0 & 0 & 0 & 0 \\	
0 & 0 & 1 & 0
\end{array}\right)
\end{equation}

Indeed, the optimal solution to Eq. (1) is given by assigning airtimes 
to the matrices in Eq. (4) only:
\begin{eqnarray}
x_1=x_2=x_3=0,x_4=1,x_5=1,x_6=1\nonumber
\end{eqnarray}

This solution requires three units of airtime, which is less than 
four units in Eq. (2).

The above example gives rise to an interesting observation: 
the matrices in Eq. (4) are the transposes of the matrices 
in Eq. (3). Thus, we have the following theorem,

\begin{theorem}
If $M^{(k)}$ conforms to constraints R1, R2, and R3 as 
defined in Definition \ref{def:mtr}, 
then so does its transpose.
\end{theorem}

\textbf{Proof:}  
It is obvious that constraints R1 and R2 are 
still fulfilled in the transposed matching 
matrix. The direction of transmission on 
a link simply gets reverse in the transposed 
matching matrix, so that the transmitters 
become the receivers, and vice versa. 
Thus, the half-duplexity constraint,
R3 is still fulfilled. 
\done

In the four-node example above, the set of 
maximal matching matrices in Eq. (1) can be 
found from the MIS in Eq. (2). More specifically, 
each MIS in Eq. (2) leads to two maximal matching 
matrices. In one matrix, each vertex in MIS is 
a transmit node and it transmits on all its outgoing links; 
in the other one, each vertex in MIS is a receive node 
and it receives on all its incoming links.
Thus, each MIS \textit{induces} two maximal matching matrices.

\begin{figure}[t]
\centering
\includegraphics[width=4.8cm]{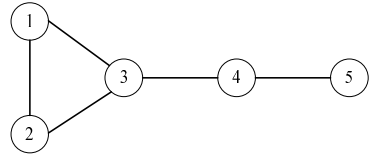}
\caption{A five-node network}
\label{fig:5node-ex}
\end{figure}

A question then is whether all maximal matching 
matrices are induced from an MIS. Unfortunately, 
the answer is no. Consider the following five-node network, 
as shown in Fig. \ref{fig:5node-ex}. A possible maximal 
matching matrix to this network is shown as follows:
\begin{displaymath}
M'=
\left(\begin{array}{ccccc}
0 & 0 & 1 & 0 & 0\\
0 & 0 & 1 & 0 & 0\\
0 & 0 & 0 & 0 & 0\\
0 & 0 & 0 & 0 & 0\\	
0 & 0 & 0 & 1 & 0
\end{array}\right)
\end{displaymath}

\begin{figure}[t]
\centering
\includegraphics[width=4.8cm]{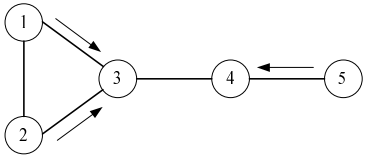}
\caption{The active links in the five-node network}
\label{fig:5node-ex-a}
\end{figure}

In this matching, nodes 1, 2, and 5 are transmitters; 
and node 3 and 4 are receivers. The transmitters do 
not form an MIS because nodes 1 and 2 are neighbors. 
The receivers do not form an MIS because nodes 3 
and 4 are neighbors. Fig. \ref{fig:5node-ex-a} 
depicts the active links in 
maximal matching matrix $M'$.

\begin{figure}[t]
\centering
\includegraphics[width=4.8cm]{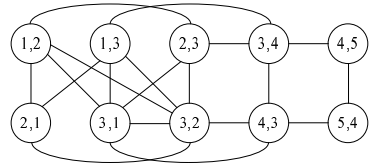}
\caption{The conflict graph for the five-node network}
\label{fig:5node-CG}
\end{figure}

In general, the number of maximal matchings 
can be more than twice the number of MIS. 
Then, how are maximal matching matrices related 
to MIS? To establish the relation, we need to 
model the network with a different graph. 
We use a \textit{conflict} graph to describe 
the relationship between two 
conflicting links. In this graph, 
each directional link is denoted 
by a vertex, and there is an edge between two
vertices if the two associated links cannot 
be active at the same time. 
The conflict graph for the five-node 
network above is shown as Fig. \ref{fig:5node-CG}, 
where vertex $(i,j)$ represents link $l_{ij}$.
With the modified graph, we can then 
formulate the problem in Eq. (1). 
In the traffic vector, $t_v=t_{ij}$ 
where $v$ represents the vertex 
corresponding to link $l_{ij}$.

\subsection{Problem Restatement}
In optimization problem defined in Eq. (1), we represent a matching by a matrix $M^{(k)}$ for pedagogical purposes. We now define a more economical representation.

\begin{definition}
A matching $A$ in an MTR network is a subset of links that conform to R1, R2 and R3.
\end{definition}

\begin{definition}
A matching is said to be maximal if it is not contained in any other matching.
\end{definition}

Let $E=\{E_j:1 \leq j \leq |E|\}$ be the set of all the feasible
matchings. The number of time slots allocated to
each feasible matching $E_j$ is denoted by a non-negative variable $u_j$.

Let $N$ be the total number of links in the network. We introduce an $N\times|E|$ incidence matrix $Q$ with elements $q_{ij}$ such that 
\begin{displaymath}
q_{ij}=\left\{ \begin{array}{ll}
1, & \textrm{if link $i$ is in matching $E_j$,}\\
0, & \textrm{otherwise.}
\end{array} \right.
\end{displaymath}
where each column in $Q$ indicates the links in a matching.

We also convert the traffic matrix $T=[T_{ij}]$ to a vector $\mathbf{f}=(f_{ij})^T$, 
where $f_{ij}=T_{ij}$  for $i$, $j$ such that $T_{ij}\neq 0$. 
Then, the problem defined in Eq. (\ref{eqn:central}) can be casted as a linear program as follows:
\begin{eqnarray}
\label{eqn:lp}
\min & \mathbf{e}^T\mathbf{u}  \nonumber\\
s.t. & Q \cdot \mathbf{u} \geq \mathbf{f} \nonumber\\
& \mathbf{u} \geq 0 \textrm{ for all } k
\end{eqnarray}
where $\mathbf{e}$ is a vector whose
components are all 1's, $\mathbf{f}=(f_1,f_2,...,f_{N})^T$ and $\mathbf{u}=(u_1,u_2,...,u_{|E|})^T$.

The difficulty of the above problem lies in how to find all matchings $Q$ (equivalent to finding all the independent sets in the associated conflict graph, which is NP-complete). This motivates us to investigate heuristic algorithms to solve this problem. We will present two computationally efficient algorithms in the next section.

\section{Heuristic Algorithms}
\label{sec:alg}
In this section, we propose two heuristic algorithms, heavy-weight-first (HWF) algorithm and max-degree-first (MDF) algorithm to solve the link-scheduling problem defined in Eq. (5). HWF is a greedy algorithm that always chooses links with the maximum traffic demand (the heaviest weight) into the scheduling set during each round until all the traffic is satisfied. MDF, on the other hand, chooses links with the maximum degree in the conflict graph during each round.

Both HWF and MDF make use of a conflict graph to capture constraints R1, R2 and R3. In the conflict graph, a link $(i,j)$ is represented by a vertex. Two links, $(i,j)$ and $(k,l)$, conflict with each other if and only if $i=l$ or $j=k$. An edge is drawn between the vertices representing $(i,j)$ and $(k,l)$ if they conflict with each other.

\subsection{Heavy-Weight-First Algorithm}

\begin{algorithm}[ht]
\footnotesize
\caption{Heavy-Weight-First Algorithm}
\label{alg:HWF}
\begin{algorithmic}[1]
\REQUIRE the network $G$, traffic demand $\mathbf{f}$
\STATE construct matching $A$ according to $G$;
\STATE generate the conflict graph $CG$;
\WHILE{the traffic demand $\mathbf{f} \neq 0$}
\STATE sort matching $A$ of the links in a descending order based on the traffic demands;
\STATE $E_i:=\emptyset$;
\STATE $m:=1$;
\WHILE{$m \neq N$}
\STATE pick the element $A(m)$;
\IF{adding $A(m)$ into $E_i$ does not cause conflict in set $E_i$}
\STATE add $A(m)$ into $E_i$;
\ENDIF
\STATE $m:=m+1$;
\ENDWHILE
\STATE $t_{\min}:=\min_{e\in E_i}{t_e}$ ($t_e \in \mathbf{f}$);
\STATE $u_i:=t_{\min}$;
\FOR{$j:=1$ to $N$}
\STATE Update the weight of every link in $E_i$ to $t_j:=t_j-t_{\min}$;
\IF{$t_j=0$}
\STATE remove link $j$ from matching $A$;
\ENDIF
\ENDFOR
\STATE output $E_i$ and $u_i$;
\STATE$i:=i+1$
\ENDWHILE
\end{algorithmic}
\end{algorithm}

In Heavy-Weight-First algorithm (HWF), we first sort the links 
according to their traffic demands in a descending order. 
To construct a matching, $E_i$, we go through the link one by one. A link will be included into $E_i$ if it does not conflict with the existing links in $E_i$  according to the conflict graph. Once we have gone through all the links in the sorted list, we then identify the link in $E_i$  with the least amount of traffic.  Let us say this is link $k$, with traffic $f_k$ . We then assign $f_k$  time slots to matching $E_i$. We subtract $f_k$ from the traffic of all the links in $E_i$, and remove link $k$ and other links in $E_i$ with the same amount of traffic (if any) from further consideration: there is not traffic left to be scheduled for these links. The links are then resorted according to their remaining traffics. The above process is iterated until all traffic demands are met. At most $N$ iterations are needed, since each iteration removes at least one link from further consideration.

\subsection{Max-Degree-First Algorithm}

\begin{algorithm}[t]
\footnotesize
\caption{Max-Degree-First Algorithm}
\label{alg:MDF}
\begin{algorithmic}[1]
\REQUIRE the network $G$, traffic demand $\mathbf{f}$
\STATE construct matching $A$ according to $G$;
\STATE generate the conflict graph $CG$;
\STATE calculate the degree of each link based on the conflict graph $CG$;
\WHILE{the traffic demand $f \neq 0$}
\STATE sort matching $A$ of the links in a descending order of the degrees;
\STATE $E_i:=\emptyset$;
\STATE $m:=1$;
\WHILE{$m \neq N$}
\STATE pick the element $A(m)$;
\IF{adding $A(m)$ into $E_i$ does not cause conflict in the set $E_i$}
\STATE add $A(m)$ into $E_i$;
\ENDIF
\STATE $m:=m+1$;
\ENDWHILE
\STATE $t_{\min}:=\min_{e\in E_i}{t_e}$ ($t_e \in \mathbf{f}$);
\STATE $u_i:=t_{\min}$;
\FOR{$j:=1$ to $N$}
\STATE Update the weight of every link in $E_i$ to $t_j:=t_j-t_{\min}$;
\IF{$t_j=0$}
\STATE remove link $j$ from matching $A$;
\STATE update the conflict graph $CG$;
\STATE update the degree of each vertex in $CG$;
\ENDIF
\ENDFOR
\STATE output $E_i$ and $u_i$;
\STATE$i:=i+1$
\ENDWHILE
\end{algorithmic}
\end{algorithm}

In MDF, we sort the links according to their degrees in the conflict graph in a descending order. Other than the different way of sorting the links, the algorithm of MDF is essentially the same as that of HWF. In particular, at least one link will be removed at the end of each iteration. The degrees of the neighbors to this link will be updated. The remaining links will also need to be resorted accordingly before the next iteration.  During each iteration, at least one link will be removed. Thus, similar to HWF, MDF needs at most $N$ iterations.

\section{Simulation Results}
\label{sec:sim}
We have conducted extensive simulation experiments on a Pentium 2.86GHz PC with 2GB memory. 
We consider several types of networks: (1) regular networks, including the linear network in Fig. \ref{fig:linear}, the grid network in Fig. \ref{fig:grid}, the ring network in Fig. \ref{fig:ring}, and the fully-connected network in Fig. \ref{fig:4node-complete}; (2) random networks with varying degrees of connectivity. We also consider various traffic demands, \textit{symmetric} as well as \textit{asymmetric}.

Symmetric traffic demands mean that $f_{ij}=f_{ji}$ for all pairs of $i$-$j$ neighbors. Symmetric traffic demands, on the other hand, mean that $f_{ij}\neq f_{ji}$ for some $i$-$j$ neighbors.
  
When the network topology is bipartite, we can easily find the optimal solution for the link scheduling problem. Details about this issue are presented in Appendix A. Note that since a tree network topology can always be organized into a bipartite graph, link scheduling in an MTR tree network is also easy.

\subsection{Performance in Regular Networks}
\begin{figure}[t]
\centering
\includegraphics[width=5.6cm]{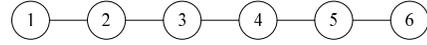}
\caption{The linear network}
\label{fig:linear}
\end{figure}

For the investigation of regular networks, we first present the simulation results of the linear network in Fig. \ref{fig:linear}. For this small network, we can easily find the optimal solution by simple hand calculation.

We present the simulation results in Table \ref{tab:linear}. The link set is $\{(1,2),(2,1),(2,3),(3,2),(3,4),(4,3),(4,5),(5,4),(5,6),(6,5)\}$.
Both HWF and MDF achieve the optimal solutions for both symmetric traffic demands (simulation No. 1 and simulation No. 2) and asymmetric traffic demands (simulation No. 3).

\begin{table}[ht]
\caption{Simulation results for the linear network}
\centering
\begin{tabular}{c|c|c|c|c}
\hline
No. & Traffic demands $\mathbf{f}$ & HWF & MDF & optimal\\
\hline
\hline
1 & $[5, 5, 5, 5, 5, 5, 5, 5, 5, 5]$ & 10 & 10 & 10 \\
\hline
2 & $[6, 6, 4, 4, 8, 8, 5, 5, 7, 7]$ & 16 & 16 & 16 \\
\hline
3 & $[6, 3, 4, 5, 7, 8, 5, 2, 7, 9]$ & 16 & 16 & 16 \\
\hline
\end{tabular}
\label{tab:linear}
\end{table}

\begin{figure}[t]
\begin{tabular}{c c c}
\begin{minipage}[t]{2.6cm} 
\centering
\includegraphics[width=2.6cm]{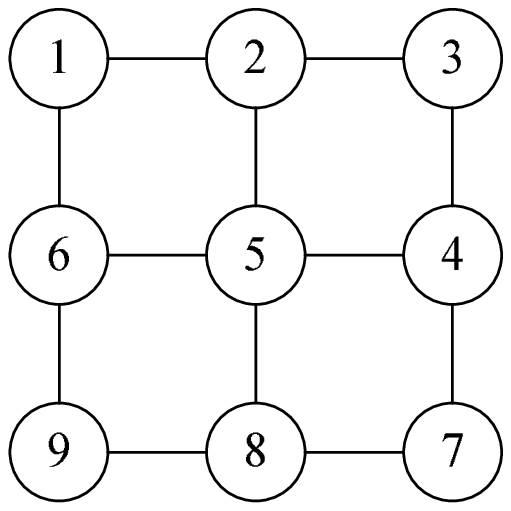}
\caption{The grid network}
\label{fig:grid}
\end{minipage}
 &
 \begin{minipage}[t]{2.7cm}
\includegraphics[width=2.7cm]{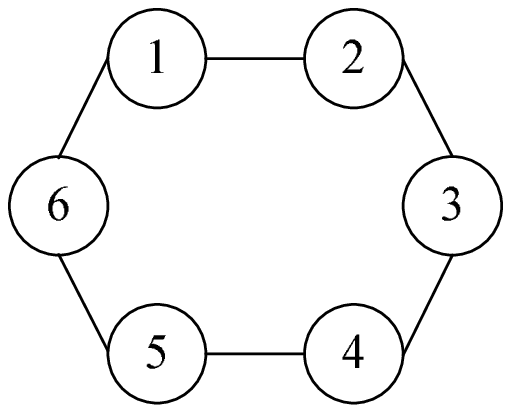}
\caption{The ring network}
\label{fig:ring}
 \end{minipage}
 &
\begin{minipage}[t]{2.6cm} 
\centering
\includegraphics[width=2.3cm]{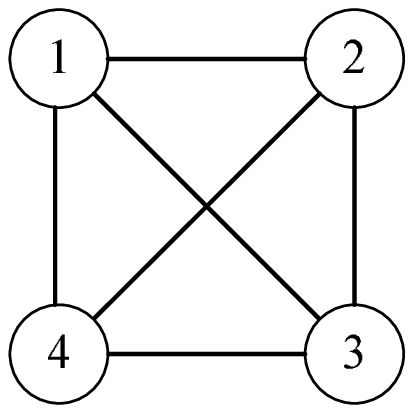}
\caption{The fully-connected network}
\label{fig:4node-complete}
\end{minipage}
 \\
\end{tabular}
\end{figure}

In the second set of simulations, we consider a grid network with nine nodes, as shown in Fig. \ref{fig:grid}. The simulations results are shown in Table. \ref{tab:grid}. The link set is $\{(1,2),(1,6),(2,1),(2,3),(2,5),(3,2),(3,4),(4,3),(4,5\\),(4,9),(5,2),(5,4),(5,6),(5,8),(6,1),(6,5),(6,7),(7,6\\),(7,8),(8,5),(8,7),(8,9),(9,4),(9,8)\}$. The simulation results show that both HWF and MDF obtain the optimal solution of 10 for the symmetric traffic demands (simulation No. 1). But for asymmetric traffic demands (simulation No. 2), HWF obtains a solution of 20, which is greater than the optimal solution of 18; while MDF obtains the optimal solution of 18.

\begin{table}[ht]
\caption{Simulation results for the grid network}
\centering
\begin{tabular}{c|c|c|c|c}
\hline
No. & Traffic demand $\mathbf{f}$ & HWF & MDF & optimal\\
\hline
\hline
1 & $[5, 5, 5, 5, 5, 5, 5, 5, 5, 5, 5, 5,$ & 10 & 10 & 10 \\
 & $5, 5, 5, 5, 5, 5, 5, 5, 5, 5, 5, 5]$ & & & \\
\hline
2 & $[7, 8, 8, 4, 7, 2, 8, 1, 3, 1, 1, 9,$ & 20 & 18 & 18 \\
  & $7, 4, 10, 1, 5, 4, 8, 8, 2, 5, 5, 7]$ & & & \\
\hline
\end{tabular}
\label{tab:grid}
\end{table}

The third set of simulations are based on a ring network with six nodes, as shown in Fig. \ref{fig:ring}. Simulations results are shown in Table \ref{tab:ring}. The link set is $\{(1,2), (1,6), (2,1), (2,3), (3,2), (3,4), (4,3), (4,5), (5,4)\\, (5,6), (6,1), (6,5)\}$. The simulation results show that both HWF and MDF achieve the optimal solution of 10 for symmetric traffic demands (simulation No. 1) and asymmetric traffic demands (simulation No. 2).

\begin{table}[ht]
\caption{Simulation results for the ring network}
\centering
\begin{tabular}{c|c|c|c|c}
\hline
No. & Traffic demands $\mathbf{f}$ & HWF & MDF & optimal\\
\hline
\hline
1 & $[5, 5, 5, 5, 5, 5, 5, 5, 5, 5, 5, 5]$ & 10 & 10 & 10 \\
\hline
2 & $[2, 5, 10, 3, 4, 6, 7, 8, 9, 11, 4, 12]$ & 23 & 23 & 23 \\
\hline
\end{tabular}
\label{tab:ring}
\end{table}

We have also conducted 1,000 simulation experiments (with different traffic demands) for each of the following networks: the linear network (Fig. \ref{fig:linear}), the grid network (Fig. \ref{fig:grid}), the ring network (Fig. \ref{fig:ring}) and the fully-connected network (Fig. \ref{fig:4node-complete}).

In order to compare the solutions obtained by the proposed algorithms with optimal solutions, we introduce the \textit{percentage cost penalty} \cite{liqun:icc2008} as a performance measure. Its definition is as follows:
\begin{equation}
P=\frac{T-T_{opt}}{T_{opt}}\times 100\%
\end{equation}
where $T$ denotes the total number of time slots obtained by the heuristic algorithm and $T_{opt}$ is the total number of time slots in the optimal solution. 

We compute $P$ of HWF and MDF over the 1,000 experiments and present the averaged $P$ values in Table 4. In each experiment, we generate a random traffic demand vector $\mathbf{f}$, where each element of $\mathbf{f}$ conforms to a discrete uniform distribution with values ranging from 1 to 10.

The results in Table \ref{tab:regular} show that MDF outperforms HWF in the linear network, the grid network and the ring network. But HWF performs better in the fully-connected network. The above results can be explained intuitively as follows. Recall that a link is removed at the end of each iteration in MDF or HWF. The nature of MDF is such that the link being removed has a high degree in the conflict graph. In this sense, MDF tends to remove many edges in the conflict graph. As a result, in a sparsely connected network (e.g., the linear network in Fig. \ref{fig:linear}, the grid network in Fig. \ref{fig:grid} and the ring network in Fig. \ref{fig:ring}), many links become conflict-free after several iterations.

\begin{figure}[t]
\centering
\includegraphics[width=5.2cm]{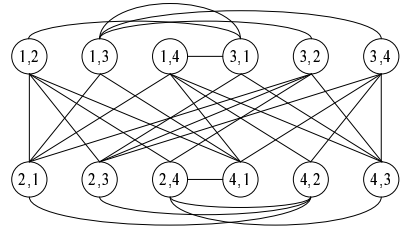}
\caption{The conflict graph for the fully-connected network}
\label{fig:4node-complete-CG}
\end{figure}

By contrast, in a densely connected network, (e.g., the fully-connected network in Fig. \ref{fig:4node-complete}), the links are so closely connected together in the conflict graph (e.g., in Fig. \ref{fig:4node-complete-CG}) that very few links can become conflict-free even after several iterations. As to why HWF tends to have better performance than MDF when the network is densely connected, we shall defer the intuitive explanation to Section 4.2, where we focus on large-scale random networks.

\begin{table}[ht]
\caption{Simulation results for random traffic in varied regular networks}
\centering
\renewcommand{\arraystretch}{1.0}
\begin{tabular}{m{2.5cm}|m{2.5cm}|m{2.5cm}}
\hline
 \centering & \centering Average cost penalty of HWF & \begin{center} Average cost penalty of MDF \end{center}\\
\hline
\hline
\centering Linear network (Fig. \ref{fig:linear})  & \centering $5.49\%$ & \begin{center} $0\%$ \end{center} \\
\hline
\centering Grid network (Fig. \ref{fig:grid}) & \centering $8.16\%$ & \begin{center} $0\%$ \end{center} \\
\hline
\centering Ring network (Fig. \ref{fig:ring}) & \centering $7.97\%$ & \begin{center} $0\%$ \end{center}\\
\hline
\centering Fully-connected network (Fig. \ref{fig:4node-complete}) & \centering $4.04\%$ & \begin{center} $9.15\%$ \end{center}\\
\hline
\end{tabular}
\label{tab:regular}
\end{table}

\subsection{Performance in Random Networks}

For comparison purposes, we carry out \textit{exhaustive search} to find optimal solutions. We compare the average runtime of HWF and MDF with that of exhaustive search. To reduce the runtime of exhaustive search, we use a branch-and-bound algorithm, first proposed in \cite{land:1960}. 

We generate random network topologies, represented by random matrix $G$, which is symmetric with zero diagonal. In $G$, entry $g_{ij}=1$ if there is a pair directional links between nodes $i$ and $j$; and $g_{ij}=0$ otherwise. For our simulation experiments, $\Pr[g_{ij}=1]=p=0.5$, $\forall i,j$. Thus, the networks being simulated are densely connected in that a node is connected to half of the other nodes on average. The number of nodes in the simulated networks is $n=6$. Thus, the maximum number of unidirectional links is $n(n-1)=30$. Each directional link will conflict with at most $2n-1$ other links under the MTR constraints. Thus, in the associated conflict graph, each link has a degree ranging from 1 to 11 when $n=6$.

\begin{figure*}[ht]
\begin{tabular}{c c c}
\begin{minipage}[t]{5.4cm} 
\centering
\includegraphics[width=5.0cm]{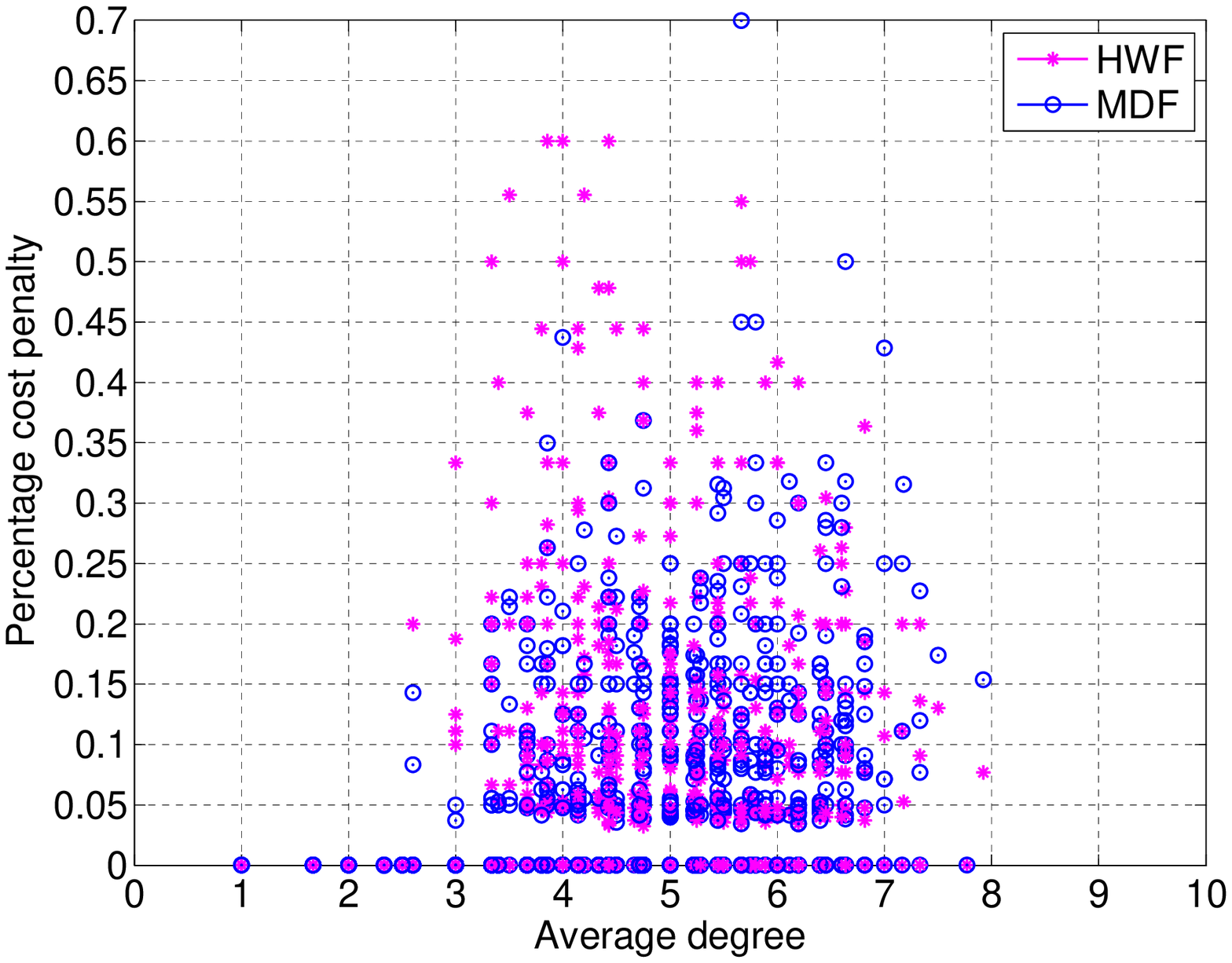}
\caption{The percentage cost penalty: symmetric traffic}
\label{fig:random1}
\end{minipage}
 &
 \begin{minipage}[t]{5.4cm}
\includegraphics[width=5.0cm]{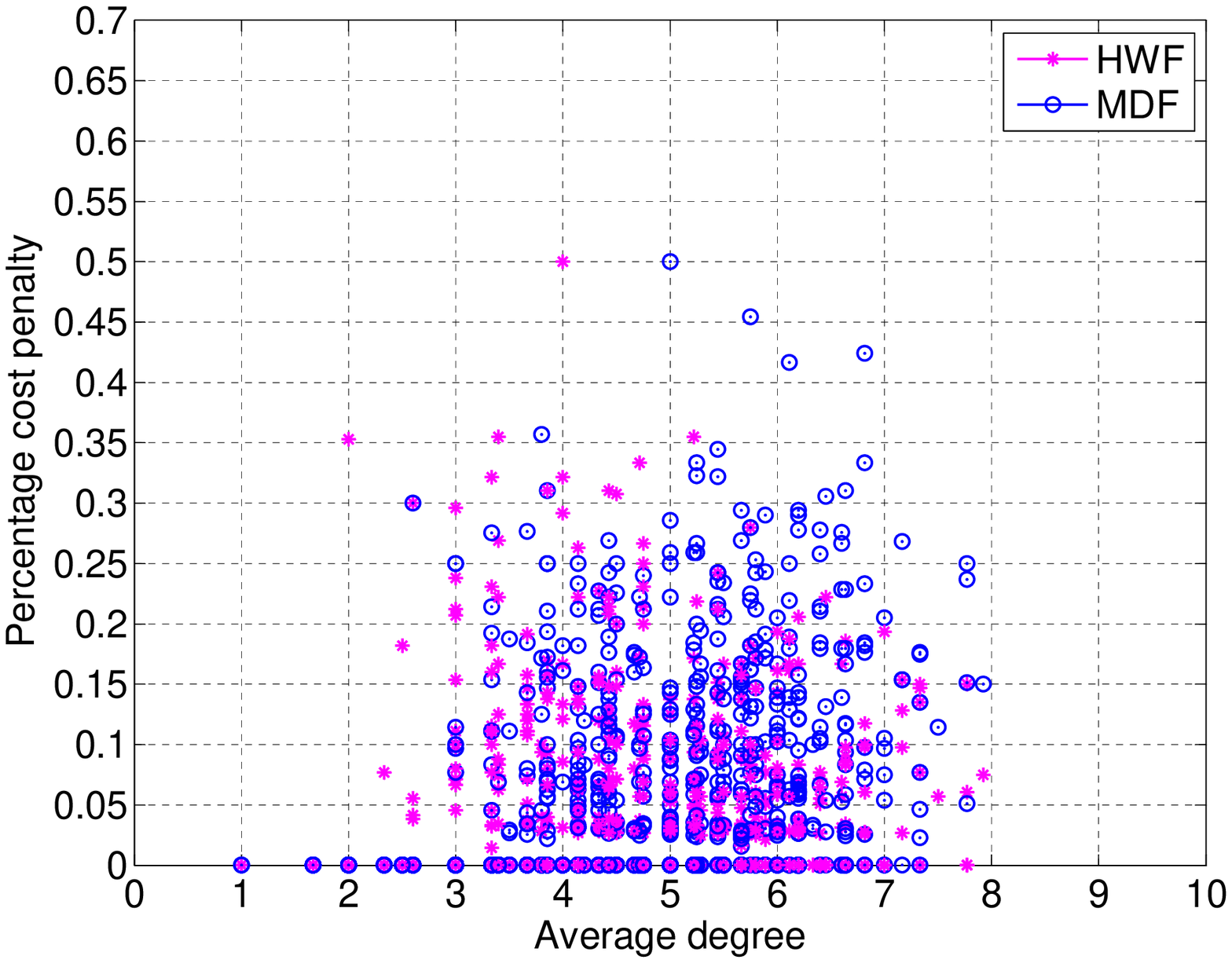}
\caption{The percentage cost penalty: asymmetric traffic}
\label{fig:random2}
 \end{minipage}
 &
\begin{minipage}[t]{5.4cm} 
\centering
\includegraphics[width=5.0cm]{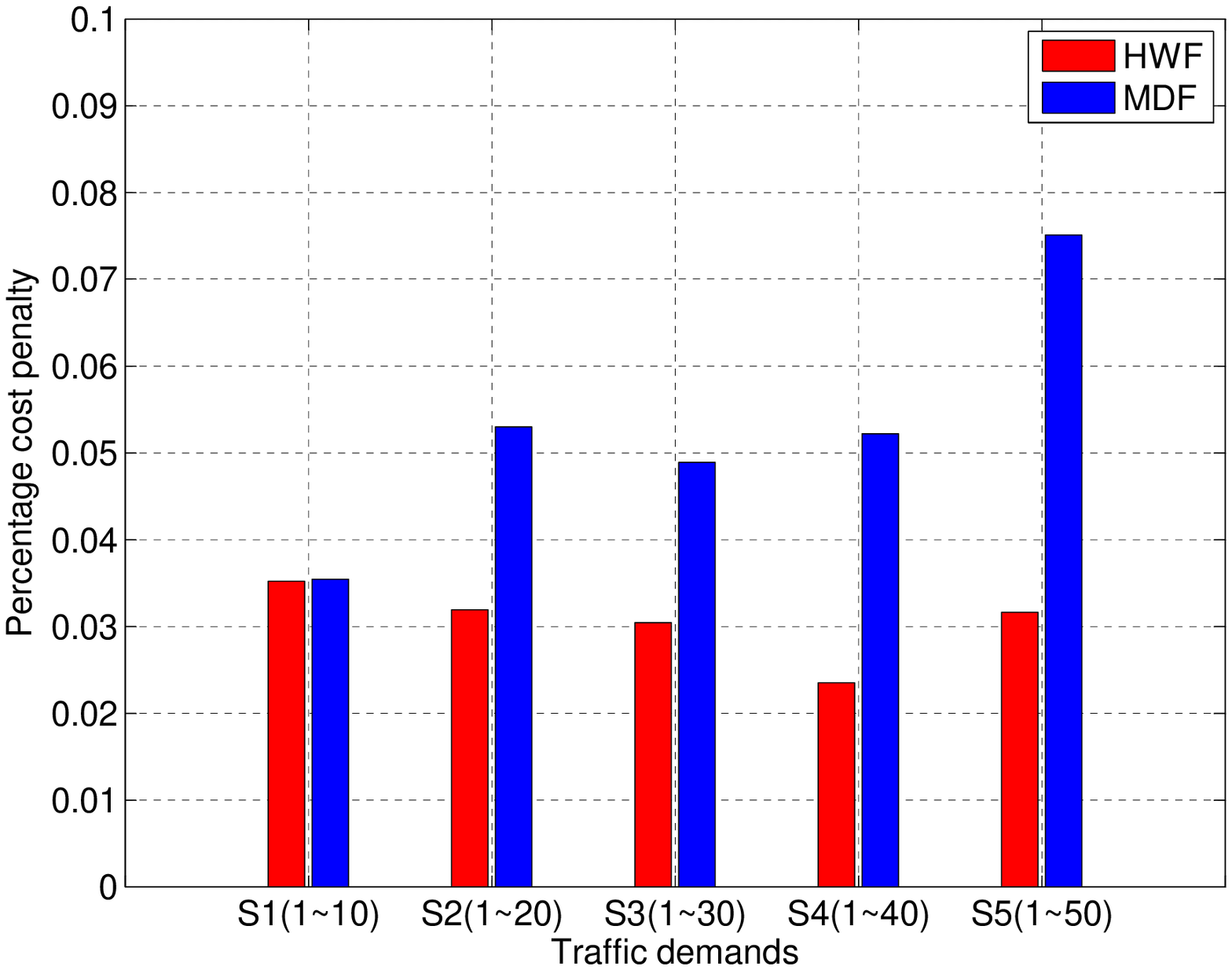}
\caption{The percentage cost penalty vs traffic demands}
\label{fig:hm}
\end{minipage}
 \\
\end{tabular}
\end{figure*}

In the first set of simulations, we consider symmetric traffic demands. If there is a link between node $i$ and node $j$ (i.e., $g_{ij}=1$), then the traffic between them, $f_{ij}=f_{ji}$, is randomly generated according to the discrete uniform distribution with values ranging from 1 to 10. We conduct 1,000 experiments and present the results in Fig. \ref{fig:random1} and Table \ref{tab:symmetry}. Each experiment is based on one random network $G$ and one associated random demand $\mathbf{f}$. Fig. \ref{fig:random1} plots percentage cost penalty versus average link degree. Table \ref{tab:symmetry} gives the statistics of the 1,000 experiments.


Fig. \ref{fig:random1} and Table \ref{tab:symmetry} show that both HWF and MDF achieve reasonably good performance. In particular, Table \ref{tab:symmetry} shows that there are nearly 800 solutions obtained by HWF and MDF with penalty cost no greater than 10\%. On average, HWF and MDF have average $P$ of 6.40\% and 5.59\%, respectively. Table \ref{tab:symmetry} also shows that the average runtime of the two algorithms is much smaller than that of exhaustive search.

\begin{table}[t]
\caption{Simulation results for symmetric traffic over random networks}
\centering
\renewcommand{\arraystretch}{0.5}
\begin{tabular}{m{1.1cm}|m{1.9cm}|m{1.8cm}|m{1cm}|m{1cm}}
\hline
 & \begin{center} No. of obtained solutions within 100\% optimality \end{center} & \begin{center} No. of obtained solutions with $P$ within 10\% \end{center} & \begin{center} Average $P$ \end{center} & \begin{center} Average runtime (second) \end{center}\\
\hline
\hline
\centering Exhaustive Search & \centering 1,000 & \centering 1,000 & \centering $0\%$ & \begin{center} 1.7522 \end{center} \\
\hline
\centering HWF & \centering 540 & \centering 781 & \centering $6.40\%$ & \begin{center} 0.0015 \end{center}\\
\hline
\centering MDF & \centering 549 & \centering 786 & \centering $5.59\%$ & \begin{center} 0.0017 \end{center}\\
\hline 
\end{tabular}
\label{tab:symmetry}
\end{table}

We have also conducted 1,000 simulations based on asymmetric traffic demands. The simulation results are presented in Fig. \ref{fig:random2} and Table \ref{tab:asymmetry}. The traffic demand $f_{ij}$ of each link $(i,j)$ is randomly generated according to the discrete uniform distribution with values ranging from 1 to 10. But the traffic in the opposite direction, $f_{ji}$ is not set to $f_{ij}$; rather, it is generated anew using the same distribution. It is shown in Fig. \ref{fig:random2} and Table \ref{tab:asymmetry} that HWF outperforms MDF in this asymmetric traffic scenario. In particular, Table \ref{tab:asymmetry} shows that HWF obtain 872 solutions with $P$ less than 10\% versus 779 obtained by MDF. On average, HWF has a lower average $P$ of 3.42\% versus 5.32\% of MDF.


\begin{table}[ht]
\caption{Simulation results for asymmetric traffic over random networks}
\centering
\renewcommand{\arraystretch}{0.5}
\begin{tabular}{m{1.1cm}|m{1.9cm}|m{1.8cm}|m{1cm}|m{1cm}}
\hline
 & \begin{center} No. of obtained solutions within 100\% optimality \end{center} & \begin{center} No. of obtained solutions with $P$ within 10\% \end{center} & \begin{center} Average $P$ \end{center} & \begin{center} Average runtime (second) \end{center}\\
\hline
\hline
\centering Exhaustive Search & \centering 1,000 & \centering 1,000 & \centering $0\%$ & \begin{center} 1.8511 \end{center}\\
\hline
\centering HWF & \centering 655 & \centering 872 & \centering $3.42\%$ & \begin{center} 0.0017 \end{center}\\
\hline
\centering MDF & \centering 568 & \centering 779 & \centering $5.32\%$ & \begin{center} 0.0019 \end{center}\\
\hline 
\end{tabular}
\label{tab:asymmetry}
\end{table}

HWF outperforms MDF in the asymmetric case because HWF can ''compact traffic demands'' as it runs. By compacting traffic demands, we mean HWF can decrease the range of the traffic demands in the network after each iteration. To see this, suppose we have a traffic demand, $\mathbf{f}=[1,2,3,4,5,6,7,8,9,10]$, where the traffic ranges from 1 to 10. Suppose that in the first iteration, links with traffic demands, 8, 9, and 10 have been chosen for scheduling. Then, we have an updated demand, $\mathbf{f}=[0,1,1,2,2,3,4,5,6,7]$ after this iteration. Now the traffic demands to be scheduled have a narrower range (i.e., 1 to 7) in the future. With compact traffic, in the later iterations, more scheduled links can be removed in each iteration because they have the same traffic demands.  MDF, on the other hand, does not have such an advantage.


Additional simulations have further verified our observation. We have conducted five additional sets of simulations. Each set of simulations are based on different values of traffic demand range. The first set of simulations are based on $S_1$ ($1\scriptsize{\sim}10$), i.e., traffic demands randomly generated according to discrete uniform distribution with values ranging from 1 to 10. The second set of simulations are based on $S_2$ ($1 \scriptsize{\sim}20 $) with values ranging from 1 to 20, etc. For each set of simulations, we calculate the averaged $P$ values for HWF and MDF over 100 simulations. Fig. \ref{fig:hm} plots the $P$ values versus different traffic ranges. It is shown in Fig. \ref{fig:hm} that MDF and HWF perform comparably with $P$ of 3.54\% and 3.52\%, respectively, when traffic demands ranges from 1 to 10. However, $P$ of MDF increases quickly as the range of traffic demands increases, while HWF remains somewhat immune to such $P$ increase. 

One possible improvement for future work is to integrate the two heuristic algorithms together. In particular, we could first sort the links according to their traffic demands in a descending order. Then, we schedule the links with the heaviest weight first. When there is a tie and two links have the same weight, we choose the link with the maximum degree in the associated conflict graph for scheduling.

\section{Conclusion}
\label{sec:con}
In this paper, we have investigated MTR networks in which a node may simultaneously send to a number of other nodes; or simultaneously receive from other nodes. This capability can potentially improve the network capacity substantially. We have (i) provided a formal specification of MTR networks for a systematic study; (ii) formulated the link-scheduling problem of minimizing the airtime usage in an MTR wireless network as a linear program (LP) and demonstrated that it is NP-hard; (iii) proposed two computationally efficient heuristic algorithms to solve this LP; and (iv) presented extensive simulation results to show that both Heavy-Weight-First (HWF) and Maximum-Degree-First (MDF) algorithms achieve good optimality and runtime performance. With regard to (iv), both HWF and MDF have average percentage cost penalty less than 10\% over 1, 000 simulation experiments with different network topologies and traffic demands. The average runtime of both HWF and MDF is less than 0.01 second.


%
%

\bibliographystyle{IEEEtran}

\bibliography{mtr-ref}

\small
\appendices
\section{}

We consider the link-scheduling problem of finding the minimum frame length in an MTR network that has a bipartite graph structure. Not all networks can be casted into a bipartite structure, however. But for those network topologies that are bipartite graphs (including tree topologies), optimal scheduling is rather simple. Note that the term ''graph'' here refers to the structure of the network itself, rather than the conflict graph associated with the network. 

\begin{figure}[ht]
\centering
\includegraphics[width=4.0cm]{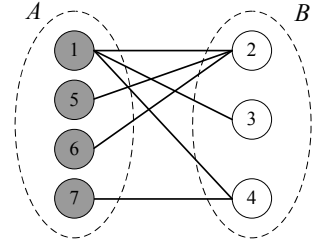}
\caption{The bipartite graph}
\label{fig:bipartite}
\end{figure}

A graph is bipartite if its vertex set can be partitioned into two subsets $A$ and $B$ so that each edge has one endpoint in $A$ and the other endpoint in $B$. Fig. \ref{fig:bipartite} shows an example of a bipartite graph.

A bipartite graph is a 2-colorable graph, i.e., we can use two colors to colorize all vertices in the graph. For example in Fig. \ref{fig:bipartite}, we can color all nodes 1, 5, 6 and 7 in $A$ gray, and all node 2, 3 and 4 in $B$ white. The gray nodes 1, 5, 6 and 7 can operate in transmit mode while the white nodes 2, 3 and 4 can operate in receive mode at the same time. Thus, MTR constraints can be easily met in a bipartite topology. Suppose the link set for the bipartite graph in Fig. \ref{fig:bipartite} is $\{(1,2),(1,3),(1,4),(2,1),(2,5),(2,6),(3,1),(4,1),(4,7)\\,(5,2),(6,2),(7,4)\}$ and the associated traffic demand vector is 
$\mathbf{f}=[1, 1, 1, 1, 1, 1, 1, 1, 1, 1, 1, 1, 1, 1]$. Then we only need two time slots to schedule all the traffic demands. 

In general, if an MTR network is bipartite, we need $T^1_{\max}+T^2_{\max}$ time slots to fulfill all the traffic demands, where $T^1_{\max}$ is the maximum traffic demand among links in one direction in the bipartite graph; and $T^2_{\max}$ is the maximum traffic demand in the other direction.

Consider the bipartite example in Fig. \ref{fig:bipartite} again. Suppose the traffic demand vector is $\mathbf{f}=[9, 8, 10, 6, 3, 4, 2, 8, 5, 7, 8, 7]$. Then, to fulfill all the traffic demands, we can use $T^1_{\max}+T^2_{\max}=10+8=18$ time slots.

We can use a simple algorithm to solve the link scheduling problem in a bipartite MTR network. In the first step, we first choose the link with the maximum traffic demand ($T^1_{\max}$) into the scheduling link set $E_1$. Then, we choose a link into link set $E_1$ if it does not conflict with the existing links in $E_1$ according the conflict graph. Repeat this process until no link can be added into link set $E_1$. We then assign $T^1_{\max}$ time slots to link set $E_1$. In the second step, we add all other remaining links into link set $E_2$. Then, we assign the maximum demand among all the remaining links, $T^2_{\max}$ time slots to link set $E_2$. Consider the above example again. The traffic demand vector is $\mathbf{f}=[9, 8, 10, 6, 3, 4, 2, 8, 5, 7, 8, 7]$. Thus, in the first step, we have link set $E_1={(1,4),(1,2), (1,3), (5,2),(6,2),(7,4)}$ and time slots  $T^1_{\max}=10$ for link $(1,4)$. In the second step, we have link set $E_2={(2,1), (2,5), (2,6), (3,1),(4,1),(4,7)}$ and time slots $T^2_{\max}=8$ for link $(4,1)$.

If the network is bipartite, we can easily solve the link scheduling problem in MTR networks. However, intentionally restricting the network topology to a bipartite graph may compromise network reliability and the network capacity \cite{Nedevschi:Mobicom2008}. Therefore, we need to consider more general topologies other than bipartite graphs. This is the motivation for our studies of more general network topologies in the main body of this paper.

\end{document}